\documentclass[12pt]{iopart}

\usepackage{graphicx,amsfonts,amsbsy, amssymb, amsthm,  mathrsfs}

\eqnobysec
\newcommand{\eqref}[1]{(\ref{#1})}

\makeatletter

\newcommand{\Rmnum}[1]{\expandafter\@slowromancap\romannumeral #1@}
\makeatother

\begin{document}
\title{Trace Formulae and Spectral Statistics for Discrete Laplacians on Regular Graphs ($\Rmnum{1}$)}
\author{Idan Oren$^{1}$, Amit Godel$^{1}$ and Uzy Smilansky$^{1,2}$}
\address{$^{1}$Department of Physics of Complex Systems,
Weizmann Institute of Science, Rehovot 76100, Israel.}
\address{$^{2}$School of Mathematics, Cardiff University, Cardiff,
Wales, UK}
 \ead{\mailto{idan.oren@weizmann.ac.il}\\
 \mailto{\ \ \ \ \ \ \ amit.godel@weizmann.ac.il}\\
 \mailto{\ \ \ \ \ \ \ uzy.smilansky@weizmann.ac.il}}

%PACS 05.45.Mt , 02.10.Ox %
\begin{abstract}
Trace formulae for $d$-regular graphs are derived and used to
express the spectral density in terms of the periodic walks on the
graphs under consideration. The trace formulae depend on a parameter
$w$  which can be tuned continuously to assign different weights to
different periodic orbit contributions. At the special value $w=1$,
the only periodic orbits which contribute are the non back-
scattering orbits, and the smooth part in the trace formula
coincides with the Kesten-McKay expression. As $w$ deviates from
unity, non vanishing weights are assigned to the periodic walks with
back-scatter, and the smooth part is modified in a consistent way.
The trace formulae presented here are the tools to be used in the
second paper in this sequence, for showing the connection between
the spectral properties of $d$-regular graphs and the theory of
random matrices.
\end{abstract}
%\ams{81Q10, 81Q50}
%\submitto{\JPA}
\section{Introduction and preliminaries}\label{sec:intro}
Discrete graphs stand at the confluence of several research
directions in physics, mathematics and computer science. Notable
physical application are e.g., the tight-binding models which are
used to investigate transport and spectral properties of mesoscopic
systems  \cite {Imry book}, and numerous applications in statistical
physics (e.g., percolation \cite {percolation}). The mathematical
literature is abundant with studies of spectral, probability and
number theory, with relation to discrete graphs \cite
{Chung,probability,AudreyBook}. Models of communication networks or
the theory of error correcting codes in computer science use graph
theory as a prime tool.

In the present series of papers, we would like to add yet another
link to the list above, namely, to study graphs as a paradigm for
Quantum Chaos. In making this contact we hope to enrich quantum
chaos by the enormous amount of knowledge accumulated in the study
of graphs, and offer the language of Quantum Chaos as a useful tool
in graph research.

The first hints of a possible connection between Quantum Chaos and
discrete graphs emerged a few years ago when Jakobson \emph{et.al.}
\cite{Rudnick} studied numerically the spectral fluctuations for
simple $d$-regular graphs (graphs where the number of neighbors of
each vertex is $d$ and no parallel or self connections are allowed).
In particular, they sampled randomly the ensemble of $d$-regular
graphs, computed the spectra of the adjacency matrices, and deduced
the mean nearest-neighbor distributions (for $d=3,4,5)$. They found
that within the statistical uncertainty, the computed distributions
match the prediction of Random Matrix Theory. The work of Terras
\cite{AudreyRev} should also be consulted in this context. The
purpose of the present work is to adopt the techniques developed in
Quantum Chaos to investigate the connection between the spectral
statistics of $d$-regular graphs and Random Matrix Theory. The main
tools which we shall use to this end are trace formulae, which, in
the present case, relate spectral statistics to the counting
statistics of periodic walks on the graphs. In the present paper,
the first in this series, we shall develop the tool kit - namely -
will derive trace formulae for regular graphs. We shall show that a
large variety of trace formulae exist, all of them provide
expression for the \emph{same} spectral density but using
\emph{differently weighted} periodic walks. We shall also show that
there exists an \emph{optimal} trace formula, in the sense that the
\emph{smooth} density coincides with the \emph{mean} density (with
respect to the $\mathcal{G}_{V,d}$ ensemble). for this optimal trace
formula, the oscillatory part stems only from a subset of periodic
orbits. These are periodic orbits in which there are no
back-scattering (reflections). In the second paper in the series, we
shall use the optimal trace formula to obtain some results which
support the conjecture that spectral statistics for regular graphs
follow the predictions of random matrix theory. In many respects,
the present study follows the development of the research in Quantum
Chaos where it was conjectured \cite{BGS} that the quantum spectra
of systems whose classical analogues are chaotic, behave
statistically as predicted by Random Matrix Theory. This conjecture,
which was originally based on a few numerical studies, brought about
a surge of research, and using the relevant (semi-classical) trace
formula \cite{Gutzwiller}, the connection with Random Matrix Theory
was theoretically established \cite{Berry, Sieber, Haake}.

To provide a proper background for the ensuing discussion, we have
to start with a short section of definitions and a summary of known
facts.

\subsection{Definitions}
A graph $\mathcal{G}$ is a set $\mathcal{V}$ of vertices connected
by a set $\mathcal{E}$ of edges. The number of vertices is denoted
by $V = |\mathcal{V}|$ and the number of edges is $E=|\mathcal{E}|$.
The $V\times V$  \emph{adjacency} (\emph{connectivity}) matrix $A$
is defined such that $A_{i,j}=s$ if the vertices $i,j$ are connected
by $s$ edges. In particular, $A_{i,i}=2s$ if there are $s$ loops
connecting the vertex $i$ to itself. A graph in which there are
loops or parallel edges, is called a \emph{multigraph}.

In the present work we mainly deal with connected \emph{simple
graphs} where there are no parallel edges ($A_{i,j}\in \{0,1\}$) or
loops ($A_{i,i}=0$). The \emph{degree} $d_i$ (\emph{valency}) is the
number of edges emanating from the vertex, $d_i=\sum_{j=1}^V
A_{i,j}$. A $d$-regular graph satisfies $d_i=d\ \ \forall\ i:  1\le
i\le V$, and for such graphs $dV$ must be even. The ensemble of all
$d$-regular graphs with $V$ vertices will be denoted by
$\mathcal{G}_{V,d}$. Averaging over this ensemble will be carried
out with uniform probability and will be denoted by $\langle
\cdots\rangle$.

To any edge $b=(i,j)$ one can assign an arbitrary direction,
resulting in two \emph{directed edges}, $e=[i,j]$ and $\hat
e=[j,i]$. Thus, the graph can be viewed as $V$ vertices connected by
edges $b=1,\cdots,E$ or by $2E$ directed edges $e=1,\cdots,2E$ (The
notation $b$ for edges and $e$ for directed edges will be kept
throughout). It is convenient to associate with each directed edge
$e =[j,i]$ its \emph{origin} $o(e) =i$ and \emph{terminus} $t(e)=j$
so that $e$ points from the vertex $i$ to the vertex $j$. The edge
$e'$ follows $e$ if $t(e)=o(e')$.

A \emph{walk} of length $t$ from the vertex $x$ to the vertex $y$ on
the graph is a sequence of successively connected vertices
$x=v_1,v_2,\cdots,v_t=y$. Alternatively, it is a sequence of $t-1$
directed edges $e_1,\cdots , e_{t-1}$ with $o(e_i)=v_i,\
t(e_i)=v_{i+1}, o(e_1)=x,\ t(e_{v-1})=y$. A \emph{closed walk} is a
walk with $x=y$. The number of walks of length $t$ between $x$ and
$y$ equals $(A^t)_{y,x}$. The graph is \emph{connected} if for any
pair of vertices there exists a $t$ such that $(A^t)_{y,x}\ne 0$.

We have to distinguish between several kinds of walks. There seems
to be no universal nomenclature, and we shall consistently use the
following:

A walk where $e_{i+1} \ne \hat{e}_i,\ ,\  1\le i\le t-2$ will be
called a \emph{walk with no back-scatter} or a \emph{nb-walk} for
short.

A walk without repeated indices will be called a \emph{path}.
Clearly, a path is a non self-intersecting nb-walk.

A \emph{t-periodic walk} is a closed walk with $t$ vertices (and $t$
edges). Any cyclic shift of the vertices on the walk produces
another t-periodic walk (which is not necessarily different from the
original one). All the t-periodic walks which are identical up to a
cyclic shift form a \emph{t-periodic orbit}. A primitive periodic
orbit is an orbit which cannot be written as a repetition of a
shorter periodic orbit.

Amongst the $t$-periodic orbits we shall distinguish those which do
not have back-scattered edges and refer to them as  periodic
nb-orbits. The frequently used term \emph{cycles}, stands for
periodic paths (non self intersecting nb-orbits).

In order to count periodic walks, it is convenient to introduce the
$2E\times2E$ matrix $B$ which describes the connectivity of the
graph in terms of its directed edges:
\begin{equation}
B_{e,e'}=\delta_{t(e),o(e')}\ . \label{eq:Bmatrix}
\end{equation}
The matrix which singles out edges connected by back-scatter is
\begin{equation}
J_{e,e'}=\delta_{\hat{e},e'}\ . \label{eq:Jmatrix}
\end{equation}
The Hashimoto connectivity matrix \cite {hashimoto89}
\begin{equation}
Y=B-J\ ,
\end{equation}
enables us to express the number of $t$ periodic nb-walks as $\tr
Y^t$. A slightly more general form:
 \begin{equation}
  Y(w)=B-wJ\ ,\  w\in \mathbb{C}\ \ \label{eq:yofw}
 \end{equation}
gives a weight $1$ to transmission and weight $1-w$ to back-scatter.
Now,  $\tr Y^t(w)=\sum_g N(t;g)(1-w)^{g}$, where $N(t;g)$ is the
number of $t$ periodic walks with exactly $g$ back-scatters.
Clearly, $\tr Y^t(w)$ can be considered as a generating function for
counting periodic walks with specific $t$ and $g$:
\begin{equation}
N(t;g)=\left . \frac{(-1)^g}{g!}\frac{\partial^g \tr
Y^t(w)}{\partial w^g}\right |_{w=1} \ .
 \label{eq:ntg}
\end{equation}

The discrete Laplacian on a graph is defined in general as
 \begin{equation}
 L\equiv-A+D,
 \label{eq:laplacian}
\end{equation}
where $A$ is the connectivity matrix, and $D$ is a diagonal matrix
with $D_{i,i} \equiv d_i$. It is a self-adjoint operator whose
spectrum consists of $V$ non negative real numbers. For $d$-regular
graphs $D$ is proportional to the unit matrix and therefore it is
sufficient to study the spectrum of the adjacency matrix $A$. This
will be the subject of the present paper.

The spectrum $\sigma(A)$ is determined as the zeros of the secular
function (characteristic polynomial)
\begin{equation}
Z_A(\mu) \equiv\det (\mu I^{(V)}-A)\ .
 \label{eq:ZsubV}
\end{equation}
Here, $\mu$ is the spectral parameter and $I^{(V)}$ is the unit
matrix in $V$ dimensions. The largest eigenvalue is $d$, and it is
simple if and only if the graph is connected. If the graph is
bipartite, $-d$ is also in the spectrum.

The \emph{spectral measure} (spectral density) is defined as
\begin{equation}
\rho(\mu) \equiv \frac{1}{V}\sum_{\mu_a\in \sigma(A)}
\delta(\mu-\mu_a)\ .
 \label{eq:def rho}
\end{equation}
The ``magnetic" Laplacian \cite {Avron} is defined by
\begin{equation}
L^{(M)}_{i,j}= - A_{i,j} {\rm e}^{i\phi_{i,j}} +d_i\delta_{i,j} \
 \ \ ,\ \ \ \phi_{i,j}=-\phi_{j,i} \ .
\end{equation}
The  phases $\phi_{i,j}$ attached to the edges $(i,j)$ play the
r\^ole of ``magnetic fluxes" (to be precise, the phases are the
analogue of gauge fields, and a sum of gauge fields over a cycle is
a magnetic flux). The Laplacian is complex hermitian, and therefore
the evolution it induces breaks time reversal symmetry. Again, for
$d$-regular graphs it suffices to study the \emph{magnetic adjacency
matrix}
\begin{equation}
A^{(M)}_{i,j}=A_{i,j} {\rm e}^{i\phi_{i,j}}\ .
 \label{eq:magA}
\end{equation}
The \emph{ensemble of random magnetic graphs} consists of the random
graph ensemble $\mathcal{G}_{V,d}$ with independently and uniformly
distributed magnetic phases $\phi_{i,j}$.

A weighted Laplacian is defined by:
\begin{equation}
L^{(W)}_{i,j}= - A_{i,j} W_{i,j} +W_{i,i}\delta_{i,j} \
 \ \ ,\ \ \ W_{i,j}=W_{j,i}\ \ \,\ \ \ W_{i,j}\in \mathbb{R} \ .
\end{equation}
Where $W_{i,j}$ are weights defined on the edges of the graph. We
shall restrict our attention to the weighted adjacency matrix:
$A^{(W)}_{i,j} = A_{i,j}W_{i,j}$ . The \emph{ensemble of random
weighted graphs} consists of the random graph ensemble
$\mathcal{G}_{V,d}$ with independently and uniformly distributed
weights  $W_{i,j}$ in the interval $|W_{i,j}|\le 1.$

\subsection{Background}
Adjacency matrices of random $d$-regular graphs, have some
remarkable spectral properties. An important discovery which marked
the starting point of the study of spectral statistics for
$d$-regular graphs, was the derivation of the mean spectral density
by Kesten \cite {Kesten} and McKay \cite{McKay}:
\begin{eqnarray} \hspace{-10mm}
\label{eq:Mckay} \rho_{KM}(\mu) = \lim_{V\rightarrow\infty}
\langle\rho(\mu)\rangle=\left\{\begin{array}{lcr}\frac{d}{2\pi}
\frac{\sqrt{4(d-1)-\mu^2}}{d^2-\mu^2} & \mbox{for} &
|\mu|\le2\sqrt{d-1}\\ \\ 0 & \mbox{for} &
|\mu|>2\sqrt{d-1}\end{array}\right. \ .
 \end{eqnarray}
The proof of this result relies on the very important property of
random $d$-regular graphs, namely, that almost surely every subgraph
of diameter less than $\log_{d-1}V$ is a tree. Counting periodic
orbits on the tree can be done explicitly, and using the close
relations between these numbers and the spectrum, one obtains (\ref
{eq:Mckay}).\\
If the entire spectrum of $A$ (except from the largest eigenvalue)
lies within the support $[-2\sqrt{d-1}, 2\sqrt{d-1}]$, the graph is
called Ramanujan (For a review, see e.g., \cite{RamMurty} and
references cited therein).

Trace formulae for regular graphs were discussed in the literature
in various contexts. The late Robert Brooks \cite {Brooks91} studied
the connection between the number of closed paths on a graph and its
spectrum, and proposed a Selberg-like trace formula. Some of the
ideas developed in the present work are related to Brook's results.
In this paper we derive two families of trace formulae which depend
on a continuous (complex) parameter $w$. The optimal trace formula
alluded to in the introduction is obtained when $w=1$. In this case
the trace formula uses the subset of nb-walks which have some
advantages (see for example \cite{Alon,Harrison}). We shall see that
with this choice of $w$, we get the Kesten-McKay measure as a
limiting distribution, and an oscillatory part which vanishes under
an ensemble average. We will make use of this part in the following
paper in this series of papers. The trace formula for this choice of
$w$ coincides with a trace formula derived by P. Mn\"{e}v,
(\cite{Mnev}) in an entirely different way.

A different, more physical approach, in which a scattering formalism
was used, resulted in a trace formula \cite {US07} (see also
\cite{Mizuno08}), which is formally similar to the trace formula for
the spectrum of the Laplacian on metric graphs \cite {Roth, KS}. For
$d$-regular graph it reads
\begin{equation}
\hspace{-15mm}
 \rho(\mu)= \frac{1}{\pi}
\ \frac{d}{\mu^2+ d ^2} \ + \ \frac{1}{V
\pi}\lim_{\epsilon\rightarrow 0^+} {\mathcal Im}\ \frac {{\rm d}\
}{{\rm d}\mu} \sum_{t=1}^{\infty}\ \frac{1}{t}\tr \left (
U(\mu+i\epsilon)^t\right)
 \label{eq:traceformula}
\end{equation}
where $U(\mu)$, the graph evolution operator, is a $2E\times 2E$
unitary matrix defined as
\begin{equation}
U(\mu) = i \left [-\frac{2}{d-i\mu} Y + (1-\frac{2}{d-i\mu})J \right
] = i \left [-\frac{2}{d-i\mu} B + J \right ] \ .
  \label{eq:evolution}
\end{equation}
$Y,J$ and $B$ were introduced before. The infinite sum in
(\ref{eq:traceformula}) can be written as a sum over $t$-periodic
walks. Denoting by $g$ the number of back-scattering along the walk,
\begin{equation}
\hspace{-10mm}\tr U^t = \frac{2^t }{ (d^2+\mu^2)^{\frac{t}{2}}}e^{i
t (\arctan\frac{\mu}{d}-\frac{\pi}{2})}\sum_g N(t;g) \  \frac
{\left((d-2)^2+\mu^2\right)^{\frac {g}{2}}}{2^g}
e^{-ig\arctan\frac{\mu}{d-2}}\ .
 \label{eq:tgorbits}
\end{equation}
The first term in the trace formula, (as usual, an algebraic
function of the spectral parameter),  is  referred to as the
``smooth" spectral density. It consists here of a Lorentzian of
width $d$ centered at $0$. The sum over the periodic walks is formal
and it diverges at the spectrum of the adjacency matrix. We shall
show below that there exists another continuum of trace formula and
that (\ref{eq:traceformula}) is obtained as a special case.

It was already observed in \cite {US07} that the trace formula above
is not satisfactory in the sense that the leading Lorentzian is very
different from the asymptotic {\it mean} spectral density
(\ref{eq:Mckay}). It was suggested that a re-summation of the
infinite sum would extract the difference between the Lorentzian and
the Kesten-McKay expression. Indeed, summing up the contribution of
the shortest, t=2 periodic orbits, one obtained a correction term
which exactly canceled the $\frac{1}{\mu^2}$ tails of the
Lorentzian. However, a systematic re-summation of (\ref
{eq:traceformula}) to extract $\rho_{MK}(\mu)$ requires a new
approach which will be presented in section (\ref{sec:tf}). The
discussion of this problem leads naturally to a more general
question: Can one distinguish amongst the periodic orbits on the
graph distinct subsets, each responsible to a different feature in
the spectral density? The Kesten-McKay theory seems to favor an
affirmative answer, since the asymptotic density is derived from
tree like periodic orbits. But what about the rest? Does one really
need all the legitimate periodic orbits on the graph, or can one do
with a subset? Is this distinction unique? These, and other
questions, will be further discussed in the following sections.

The paper is organized as follows. The Bartholdi identity is the
corner-stone of the theory presented in the present paper. However,
since the derivation is strictly technical, we leave it to an
appendix. The appendix reviews shortly the Bartholdi identity and
its proof, which is generalized here to include the cases of
magnetic adjacency matrices, multigraphs and weighted graphs as
well.\\
In the first section, we derive the family of trace formulae, which
depend on real $w$. Then, we
specialize to the choice $w=1$ both in the magnetic and non-magnetic cases.\\
The different forms of the trace formula will help to elucidate the
question posed in the previous paragraph regarding the different
r\^oles played by periodic orbits of different topologies.
Furthermore, the dependence on $w$ can be exploited to unravel
combinatorial information about the graph at hand. Finally, the
trace formula (\ref {eq:traceformula}) will be shown to be but one
of a (continuous) family of trace formulae which have one feature in
common, namely that they are based on traces of unitary operators of
the type (\ref {eq:evolution}). This is achieved by allowing $w$ to
assume complex values.

\section{A continuous family of Trace Formulae}
 \label{sec:tf}
In the present section we shall derive a continuous family of trace
formulae which depend on a real parameter $w$. The parameter $w$
controls the weights which are given to periodic walks with
different numbers of back-scatters. Consider the matrix $Y(w)$
(\ref{eq:yofw}). For $d$-regular graphs, the Bartholdi identity
reads (see Appendix A):
\begin{equation}
\hspace{-15mm}  \det(I^{(2E)}-s(B-wJ))=
 (1-w^2s^2)^{E-V}\det(I^{(V)}(1+w(d-w)s^2)-sA)\ .
 \label{eq:bartholdi}
 \end{equation}
 Its importance in the present context comes from the fact that it
connects the spectrum of the adjacency matrix $A$ with that of the
matrices $Y(w)=B-wJ$, which can be used to count various types of
cycles and walks on the corresponding graph. It implies that the
spectrum of $Y(w)=B-wJ$ is
\begin{eqnarray}
  \label{eq:specY-w}
\hspace{-10mm} \sigma(Y(w)) &=& \{(d-w), w,\ +w\times (E-V),
-w\times (E-V),  \nonumber \\
& & ( \sqrt{w(d-w)}\ {\rm e}^{i\phi_k},\sqrt{w(d-w)}\
{\rm e}^{-i\phi_k},\ k=1,\cdots (V-1)) \} \\
& & {\rm where} \ \ \ \phi_k=\arccos\frac{\mu_k}{2\sqrt{w(d-w)}}\
{\rm for \ all\ } \  k=1,\ldots, V-1.\nonumber
\end{eqnarray}
The $\mu_k$'s with $k =1,\cdots (V-1)$ are the non trivial
eigenvalues of the adjacency matrix, whose spectrum is ordered as a
non increasing sequence
\begin{equation}
d =\mu_{0} > \mu_{1}\ge\mu_{2}\ge \cdots \ge\mu_{V-1} \ge -d \ .
\end{equation}
($\mu_{V-1}$ assumes the value $-d$ if and only if the graph is
bipartite. The bipartite graphs are rare in $\mathcal{G}_{V,d}$ and
are excluded from the discussion from now on). If $w$ is in the
interval $[1, \frac{d-1}{2}]$, we have $2\sqrt{d-1} \le
2\sqrt{w(d-w)}\le \sqrt{d^2-1}$. This ensures that for generic
graphs one can always find a value of $w\in[1, \frac{d-1}{2}]$ so
that for all $k$, $|\mu_k|<2\sqrt{w(d-w)}$, and all the $\phi_k$'s
are real. The freedom to choose $w$ allows us to use the trace
formula for almost all $d$-regular graph, and in particular for
non-Ramanujan graphs.

 It is convenient to introduce the quantities $y_t(w)$,
\begin{equation}
y_t(w) = \frac{1}{V} \frac{\tr Y^t(w)-(d-w)^t}{(\sqrt{w(d-w)})^t}\ .
\end{equation}
which (unlike $\tr Y^t(w)$) are bounded as $t\rightarrow \infty$.
The explicit expressions for the eigenvalues of $Y(w)$ are used now
to write,
\begin{eqnarray}
\label{eq:ywt}
 \hspace{-25mm}y_t(w) =
\frac{1}{V}\left(\frac{w}{d-w}\right)^\frac{t}{2}+ \frac{d-2}{2}
\left(\frac{w}{d-w}\right)^\frac{t}{2}
(1+(-1)^t)+\frac{2}{V}\sum_{k=1}^{V-1}T_t(\frac{\mu_k}{2\sqrt{w(d-w)}})
\end{eqnarray}
where $T_t(x) \equiv \cos{(t\arccos{x})}$ are the Chebyshev
polynomials of the first kind of order $t$. The fact that the
$y_t(w)$ are bounded, is guaranteed since $\frac{w}{d-w}<1$ whenever
$w\in[1, \frac{d-1}{2}],\ {\rm and}\ d\ge 3$, and since the
Chebyshev polynomials are bounded.

Multiplying both sides of (\ref{eq:ywt}) by
$\frac{1}{\pi(1+\delta_{t,0})}T_t(\frac{\mu}{2\sqrt{w(d-w)}})$, and
summing over $t$, we get:
\begin{eqnarray}
\hspace{-25mm}
\frac{1}{\pi}\sum_{t=0}^\infty\frac{1}{(1+\delta_{t,0})}T_t(\frac{\mu}{2\sqrt{w(d-w)}})y_t(w)=
\nonumber \\
\hspace{-15mm} \frac{1}{\pi
V}\sum_{t=0}^\infty\frac{1}{(1+\delta_{t,0})}\left(\sqrt{\frac{w}{d-w}}\right)^t
T_t(\frac{\mu}{2\sqrt{w(d-w)}})+\nonumber\\
\hspace{-15mm} \frac{(d-2)}{2\pi}\sum_{t=0}^\infty
\frac{1}{(1+\delta_{t,0})}\left[\left(\sqrt{\frac{w}{d-w}}\right)^t+
\left(-\sqrt{\frac{w}{d-w}}\right)^t\right]
T_t(\frac{\mu}{2\sqrt{w(d-w)}})+\nonumber \\
\hspace{-15mm} \frac{1}{V}\sum_{k=1}^{V-1}\delta_T \left(
\frac{\mu}{2\sqrt{w(d-w)}},\frac{\mu_k}{2\sqrt{w(d-w)}} \right)
\end{eqnarray}
Where $\delta_T(x,y)$ is defined by:
\begin{equation}
 \label{eq:deltaf}
 \delta_T(x,y)\equiv \frac{2}{\pi}\sum_{t=0}^\infty
\frac{1}{1+\delta_{t,0}}T_t(x)T_t(y)\ .
\end{equation}
$\delta_T(x,y)$ is the unit operator in the $L^2[-1,1]$ space where
the scaler product is defined with a weight $\frac{1}{\sqrt{
1-x^2}}$. Indeed,
\begin{equation}
 \int_{-1}^1 \frac{dx}{\sqrt{1-x^2}}\delta_T(x,y)f(x) = f(y)\ .
\end{equation}
For $t=0,1,2$ one can easily show that $$y_0 = d-\frac{1}{V}\ ;\ y_1
= \frac{-1}{V}\sqrt{\frac{d-w}{w}}\ ;\  y_2 =
\frac{d(1-w)^2}{w(d-w)}-\frac{1}{V}\frac{d-w}{w}\ .$$ Writing $
 \sqrt{1-x^2} \cdot\delta(x-y) = \delta_T(x,y)$, and using elementary
identities involving the Chebyshev polynomials, we get an expression
for the density of states which is supported in the interval
$|\mu|\le 2\sqrt{w(d-w)}$ :
\begin{equation}
\rho(\mu) =
\rho^{smooth}(\mu;w)+\rho^{osc}(\mu;w)+\frac{1}{V}\rho^{corr}(\mu;w)\
,
  \label{eq:density-w}
\end{equation}
where
\begin{eqnarray}
\hspace{-25mm}
\rho^{smooth}(\mu;w) &=&\nonumber \\
\hspace{-25mm}& &\frac{d/(2\pi)}{\sqrt{4w(d-w)-\mu^2}}\left(
1-\frac{(d-2w)(d-2)}{d^2-\mu^2}+\frac{(w-1)^2(\mu^2-2w(d-w))}{w^2(d-w)^2}
\right)\nonumber \\
\hspace{-25mm} \rho^{osc}(\mu;w) &=&  \frac{1}{\pi}
\sum_{t=3}^\infty \frac{y_t(w)}{\sqrt{4w(d-w)-\mu^2}}\  T_t \left(
\frac{\mu}{2\sqrt{w(d-w)}}\right)
\nonumber\\
\hspace{-25mm}&=& \frac{1}{\pi}\mathcal{R}e\left ( \sum_{t=3}^\infty
\frac{y_t(w)}{\sqrt{4w(d-w)-\mu^2}}\ {\exp}^{i t\arccos\left(
\frac{\mu}{2\sqrt{w(d-w)}}\right) } \right )
\nonumber\\
\hspace{-25mm} \rho^{corr}(\mu;w)
&=&\frac{-1}{2\pi\sqrt{4w(d-w)-\mu^2}} \left(1+
\frac{\mu}{w}+\frac{\mu^2-2w(d-w)}{w^2}+\frac{d-2w}{d-\mu} \right) \
.
  \label{eq:density-w specific}
\end{eqnarray}
\vspace{5mm}
 \begin{figure}[!h]
  \centering
 \scalebox{0.5}{\includegraphics{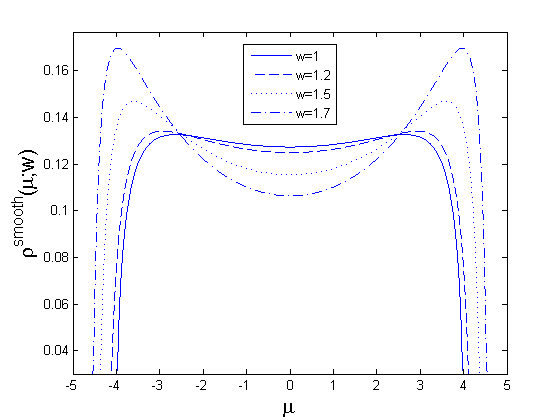}}
 \caption{$\rho^{smooth}(\mu;w)$ for $5$ regular graphs for various values of $w$:\\
  solid: $w=1$, dashed: $w=1.2$, dotted: $w=1.5$, dash-dot:
  $w=1.7$.}
  \label{fig:density of various w}
\end{figure}
Equations (\ref{eq:density-w},\ref{eq:density-w specific}) are the
main result of this section. In form they are very similar to well
known trace formulae from other branches of Mathematical Physics.
They are composed of a smooth part $\rho^{smooth}(\mu;w)$ which is
an algebraic expression in $\mu$ as shown in Figure
(\ref{fig:density of various w}), and an oscillatory part, computed
from information about the periodic walks. The amplitudes are
obtained from the (properly regularized) count of $t$-periodic
walks, and the phase factors explicitly given in the second line,
are the analogues of the ``classical actions" accumulated along the
walks. (\ref{eq:density-w}) is a generalization of
(\ref{eq:traceformula}), as will be shown in (\ref{sec:unitary})

It is important to notice that the left hand side of
(\ref{eq:density-w}) does not depend on $w$, because the density of
states of the adjacency operator depends only on the graph and not
on the choice of $w$. Therefore, the right hand side must also be
$w$-independent. In other words, the $w$ dependence of the smooth
part, is offset by a partial sum of the oscillatory part. This is
reminiscent of the partition of the trace formula
(\ref{eq:traceformula}), where the smooth part is a Lorentzian, and
it was shown that the sum over 2-periodic orbits gives a
contribution which exactly cancels the leading $1/\mu^2$ behavior
for large $|\mu|$.

Having the freedom to choose $w$, it is natural to look for the most
appropriate or convenient partition of the spectral density into
smooth and oscillatory parts. We shall show in the sequel that this
is obtained when $w=1$, because the \emph{smooth} density coincides
with the \emph{mean} density (with respect to the
$\mathcal{G}_{V,d}$ ensemble) (\ref {eq:Mckay}).

Finally, we also mention that a trace formulae for multigraphs can
be derived in an analogous fashion. The smooth part remains
unaltered, and in the oscillatory part, the sum starts from $t=1$
since loops are allowed.

\subsection{Trace formula in terms of periodic nb-walks ($w=1$)}

The case $w=1$ plays a special r\^ole in the present theory and its
applications. The fact that the smooth part of the trace formula is
identical to the Kesten-McKay expression was mentioned above, and
will be discussed further in the sequel. As will be shown, this is a
direct consequence of the fact that the counting statistics of
$t$-periodic nb-walks in the $\mathcal{G}_{V,d}$ is Poissonian for
$t < \log_{d-1}V$ with $\left \langle \tr Y^t \right \rangle =
(d-1)^t$ \cite {Balobas}.

The trace formula can be obtained by substituting $w=1$ in
(\ref{eq:density-w specific}). Alternatively, one can start from the
Bass formula for $d$-regular graphs \cite {bass}:
\begin{equation}
 \det(I^{(2E)}-sY)=
 (1- s^2)^{E-V}\det(I^{(V)}(1+ (d-1)s^2)-sA)\ .
 \label{eq:bass}
 \end{equation}
where $Y$ is the Hashimoto matrix defined in section
(\ref{sec:intro}), and  $\tr Y^t$ counts the number of $t$-periodic
nb-walks. One can now follow the same steps as in the previous
section. However, this requires restricting the discussion to
Ramanujan graphs only. Under this condition, $|\mu_{k}|\le
2\sqrt{d-1},\ k = 1,\cdots (V-1) $, and the spectrum of $Y$ is
\begin{eqnarray}
   \label{eq:specY}
 \sigma(Y) &=&\left \{(d-1), 1,\ +1\times (E-V),\
-1\times (E-V), \right .\nonumber \\
& & \left . ( \sqrt{d-1}\ {\rm e}^{i\phi_k},\ \sqrt{d-1}\
{\rm e}^{-i\phi_k},\ k =1,\cdots (V-1))\right \} \\
& & {\rm where} \ \ \ \phi_k=\arccos\frac{\mu_k}{2\sqrt{d-1}}\ , \ \
\ \  0\le \ \phi_k\ \le \pi . \nonumber
 \end{eqnarray}
For large $t$, the number of $t$-periodic walks is dominated by the
largest eigenvalue, so that asymptotically $\tr Y^t \sim (d-1)^t$.
Shortening the notation by using $y_t=y_t(w=1)$ we have
\begin{equation}
y_t= \frac{1}{V} \frac{\tr Y^t-(d-1)^t}{(\sqrt{d-1})^t}
  \label{eq:y_t}
\end{equation}
which is the properly regularized number of $t$-periodic nb-walks.
Going through exactly the same steps as in section (\ref{sec:tf}),
we get:
\begin{equation}
\hspace{-25mm} \rho(\mu) = \frac{d}{2\pi}\cdot
\frac{\sqrt{4(d-1)-\mu^2}}{d^2-\mu^2}+\frac{1}{\pi}Re\sum_{t=3}^\infty
\frac{y_t}{\sqrt{4(d-1)-\mu^2}} e^{ it \arccos \left(
\frac{\mu}{2\sqrt{(d-1)}}\right)}+\frac{1}{V}\rho^{corr}(\mu)\ .
   \label{eq:trace formula}
\end{equation}
The first term is the smooth part, and can be identified as the
Kesten-McKay density. For large $t$, since $\tr Y^t$ is dominated by
$(d-1)^t$, it is clear that in the limit of large $V$, $y_t$ tends
to zero. The counting statistics of $t$-periodic nb-walks with $t <
\log_{d-1} V $ is Poissonian, with $\left \langle \tr Y^t \right
\rangle = (d-1)^t$. Thus, the mean value of $y_t$ vanishes as
$\mathcal{O}\left(\frac{1}{V}\right )$. Hence,
\begin{equation}
\lim_{V\rightarrow \infty} \left \langle \rho(\mu)\right \rangle =
\rho_{KM}(\mu).
\end{equation}
The above can be considered as an independent proof of the
Kesten-McKay formula (\ref{eq:Mckay}). The original derivation
relied on the fact that $d$-regular graphs look locally like trees,
for which the spectral density is of the form (\ref{eq:Mckay}).
Here, it emerged without directly invoking the tree approximation,
rather, it appears as a result of an algebraic manipulation.

The last term, $\frac{1}{V}\rho^{corr}(\mu)$, is the correction to
the smooth part, for finite $V$, which results from $y_0, y_1, y_2$.
It is given explicitly by
\begin{equation}
\rho^{corr}(\mu) = \frac{-1}{2\pi\sqrt{4(d-1)-\mu^2}} \left(1+
\mu+\mu^2-2(d-1)+\frac{d-2}{d-\mu} \right)
\end{equation}
Equation (\ref{eq:trace formula}) is identical to a trace formula
derived by P. Mn\"{e}v (\cite{Mnev}). Our derivation, however,
follows an entirely different path which allows us to place the
trace formula as a special case in a more general setting, and
provides the leading order correction,
$\frac{1}{V}\rho^{corr}(\mu)$.

If the graph at hand is non-Ramanujan, the theory should be modified
since some of the phases $\phi_k$ in (\ref{eq:specY}) become
complex. Consequently, the $y_t$ diverge exponentially with $t$ and
the resulting trace formula (\ref{eq:trace formula}) is ill defined.
There are two ways to circumvent this problem. Either to use the
trace formula with $w>1$, at the cost of using periodic walks with
back-scatter, or to remain with nb-walks but reformulate the trace
formula so that it describes a coarse-grained version of the
spectral density. We shall discuss below the latter option.

Consider a non-Ramanujan graph $\mathcal{G}$ and denote by
$\gamma(\mathcal{G})$ the set of eigenvalues which lie outside the
Kesten-McKay support $[-2\sqrt{d-1},2\sqrt{d-1}]$. The rigorous and
numerical information available to date on the set of eigenvalues
$\gamma(\mathcal{G})$, suggests that generic $(V,d)$ regular graphs
have the following properties \cite{Hoory,SashaSodin}.
\begin{itemize}
\item The distance between $\gamma(\mathcal{G})$ and the
Kesten-McKay support is smaller than $a V^{-\alpha}$, with $\alpha
>0$ and a positive constant $a$. Numerical simulations show that
$\alpha \approx 2/3$.
\item The cardinality of the set
$\gamma(\mathcal{G})$ is bounded, and $\left
\langle|\gamma(\mathcal{G})|\right\rangle \ =\ \mathcal{O}(1)$.
\end{itemize}

The derivation of the trace formula for Ramanujan graphs (\ref
{eq:trace formula}) used (\ref{eq:deltaf}) which expresses the
identity operator in terms of the orthonormal set of Chebyshev
polynomials. Truncating the infinite sum in (\ref{eq:deltaf}) at
$t=t_{max}$, one obtains the identity operator in the finite space
spanned by the first $t_{max}$ Chebyshev polynomials. The resulting
sum
\begin{equation}
\tilde \delta (x,x') = \frac{2}{\pi \sqrt {1-x^2}}
\sum_{t=0}^{t_{max}}\frac{1}{1+\delta_{t,0}} T_t(x) T_t(x')
\end{equation}
displays a peak centered at $x=x'$ of width $\approx
\frac{1}{t_{max}}$.  The coarse-grained spectral density for $\mu$
in the Kesten-McKay interval is defined as
\begin{equation}
\hspace{-10mm}\tilde \rho (\mu) = \frac{2\sqrt{d-1}}
{V-|\gamma(\mathcal{G})|}\ \  \cdot \sum_{\mu_k\in
\mathcal{G}\setminus\gamma(\mathcal{G})} \tilde \delta
(\frac{\mu}{2\sqrt{d-1}},\frac{\mu_k}{2\sqrt{d-1}})\ .
\end{equation}
We return now to (\ref{eq:ywt}) with $w=1$,  and separate the sum on
the spectrum to its  Ramanujan and non-Ramanujan components:
\begin{eqnarray}
\label{eq:yw=1t}
 \hspace{-15mm}y_t &=&
\frac{1}{V}\left( d-1\right)^{-\frac{t}{2}}+\frac{d-2}{2} \left(
d-1\right)^{-\frac{t}{2}} (1+(-1)^t)
\nonumber \\
\hspace{-15mm}&+& \frac{2}{V}\sum_{\mu_k\in
\mathcal{G}\setminus\gamma(\mathcal{G})}
T_t(\frac{\mu_k}{2\sqrt{d-1}}) +\frac{2}{V}\sum_{ {\mu_k\in
\gamma(\mathcal{G})} }T_t(\frac{\mu_k}{2\sqrt{d-1}})\ .
\end{eqnarray}
Upon multiplying both sides by
$\frac{1}{\pi(1+\delta_{t,0})}T_t(\frac{\mu}{2\sqrt{d-1}})$ and
performing the sum over $t=0,\ldots,t_{max}$ we have to consider in
particular the last term since it involves the exponentially
increasing contributions from the non-Ramanujan spectral values.
Consider $\mu_k\in \gamma(\mathcal{G})$. It is bounded by $|\mu_k|
\le 2\sqrt{d-1}(1+a V^{-\alpha})$ and therefore the contribution of
the sum of the last terms in (\ref{eq:yw=1t}) is bounded by
$$\frac {2 |\gamma(\mathcal{G})|}{V}\frac{\exp (t_{max}a
V^{-\alpha/2})}{\sqrt{a}V^{-\alpha/2}}.$$ Choosing $t_{max} < \frac{
V^{\alpha/2}}{a}\log V^{1-\alpha/2}$ ensures that the sum converges
to zero as $V$ increases. Summing the other terms in (\ref
{eq:yw=1t}) results in the finite $t_{max}$ analogue of the
Kesten-McKay density, which converges exponentially quickly to the
limit expression (the correction goes to zero as $z^{t_{max}},\
z<1$). The truncated oscillatory contribution arises from the
truncated sum over the left hand side of equation (\ref {eq:yw=1t}).
Truncating the $t$ sum at $t_{max}$ implies that the trace formula
cannot resolve spectral intervals of order $\frac{1}{t_{max}}$,
which are larger than the mean spectral spacing which is of order
$\frac{1}{V}$. In many applications this does not pose a severe
problem.

\subsection{The Magnetic Case (w=1)}
  \label{sec:mtf}
The magnetic spectrum differs from the corresponding non magnetic
one in one important aspect, namely, there is no analogue to the
phenomenon that the maximal eigenvalue $\mu_0$ is identically $d$.
Rather, when averaged over the magnetic ensemble, it falls right on
the boundary of the Kesten-McKay support. This can be shown by the
following heuristic argument (A rigorous proof can be found in
\cite{SashaSodin}). Consider the magnetic edge-connectivity matrix
which excludes back-scatter: $Y^{(M)} = B^{(M)}-J$. The ensemble
mean of its maximal eigenvalue can be estimated by studying the
behavior of $\left \langle \tr \left [ \left (Y^{(M)}\right)^t\right
]\right \rangle_{\Phi}$ for large $t$, where $\langle \cdots
\rangle_{\Phi}$ denotes the average over the magnetic ensemble. The
only non vanishing contributions to $\left \langle \tr \left
[\left(Y^{(M)}\right)^t\right ]\right \rangle_{\Phi}$ come from
\emph{self-tracing} periodic nb-walks where each edge is traversed
an equal number of times in both directions. The periodic walks
which asymptotically dominate $\left \langle \tr \left [
\left(Y^{(M)}\right)^t\right ]\right \rangle_{\Phi}$ consist of two
cycles which share a single vertex, and each cycle is traversed
twice in opposite directions. The common vertex enables the
inversion of the traversal direction without back-scatter. On
average, the number of such periodic walks is of order
$t(d-1)^{\frac{t}{2}}$ hence, $\langle |\eta_0 |\rangle _{\Phi}\
\simeq \sqrt{d-1}$, where $\eta_0$ is the largest eigenvalue of
$Y^{(M)}$ in absolute magnitude. Using the connection between the
spectra of $Y^{(M)}$ and $A^{(M)}$ which is implied by the Bartholdi
identity, we deduce $\langle |\mu_0 |\rangle _{\Phi}\ \simeq
2\sqrt{d-1}$.

Starting from equation (\ref{eq:m-bartholdi}), we continue by
assuming that the graph is Ramanujan in the sense that all
eigenvalues of $A^{(M)}$, including the largest one satisfy
$|\mu|<2\sqrt{d-1}$. The spectrum of $Y^{(M)}$ consists of
\begin{eqnarray}
 \hspace{-25mm}\sigma(Y^{(M)}) &=&\left \{  +1\times (E-V),\
-1\times (E-V), \right .\nonumber \\
& & \left . ( \sqrt{d-1}\ {\rm e}^{i\phi_k},\ \sqrt{d-1}\
{\rm e}^{-i\phi_k},\ k = 1,\cdots V)\right \} \nonumber \\
\hspace{-25mm} & & {\rm where} \ \ \
\phi_k=\arccos\frac{\mu_k}{2\sqrt{(d-1)}}\ , \ \ \ \  0\le \ \phi_k\
\le \pi .
 \label{eq:specY-m}
 \end{eqnarray}
The scaled traces $y_t$ are defined as
\begin{equation}
y_t = \frac{1}{V} \frac{\tr (Y^{(M)})^t}{(\sqrt{(d-1)})^t}
\end{equation}
and we end up with a trace formula, similar to  equation
(\ref{eq:trace formula}). The smooth part is again the Kesten-McKay
density. The  oscillatory part is different, because the terms in
$y_t$  contributed by $t$-periodic walks are now decorated by
magnetic phases which are accumulated along the walks. We have shown
previously that $\left \langle \tr \left [ \left
(Y^{(M)}\right)^t\right ]\right \rangle_{\Phi} \sim
t(d-1)^{\frac{t}{2}}$. Thus, $\langle y_t \rangle_{\Phi} \rightarrow
0$ if the limits $t\rightarrow \infty$ and $V\rightarrow \infty$ are
taken such that $\frac{t}{V}\rightarrow 0$. Therefore the {\it mean}
spectral density is again the Kesten-McKay measure, as in the non
magnetic $w=1$ case.

\section{Applications of the $w$-Trace Formula}
  \label{sec:applications}

The $w$-trace formula offers a bridge between the spectral and the
combinatorial aspects of graph theory. This connection can be
exploited in order to compute combinatorial quantities, as will be
shown in the present section.

Throughout this section we shall assume the large $V$ limit of the
spectral density (\ref{eq:density-w}), and  neglect the term
$\frac{1}{V}\rho^{corr}(\mu;w)$:
\begin{eqnarray}
\rho(\mu) = \rho^{smooth}(\mu;w)+\rho^{osc}(\mu;w). \nonumber
\end{eqnarray}
The explicit expressions for $ \rho^{smooth}(\mu;w)$ and
$\rho^{osc}(\mu;w)$ are given in (\ref{eq:density-w specific}).

In the first section we have shown that  $\tr Y^t(w)$ can be
considered as a generating function for the number $N(t;g)$ of $t$
periodic walks which scatter back exactly $g$ times:
\begin{equation}
N(t;g)=\left . \frac{(-1)^g}{g!}\frac{\partial^g \tr
Y^t(w)}{\partial w^g}\right |_{w=1} \ .
\end{equation}
Here, we shall use the $w$-trace formula to compute $N(t,g=1)$
explicitly and show how expressions for higher $g$ can be obtained.
To start, we take the first derivative of (\ref{eq:density-w}) with
respect to $w$. The left hand side vanishes, since $\rho(\mu)$ does
not depend on $w$. Moreover, it can be easily checked that
$\frac{d\rho^{smooth}(\mu;w)} {dw}|_{w=1}=0$. Thus, the first
derivative of $\rho^{osc}(\mu;w)$ computed at $w=1$ must vanish
identically for any $\mu$. We shall show that the first derivative
can be written in the form $\sum_l a_lT_l(\frac{\mu}{2\sqrt{d-1}})$,
and therefore the coefficients $a_l$ must vanish. This requirement
provides a recurrence relation from which $N(t,g=1)$ can be computed
for any $t$.

From now on we shall denote $\frac{d}{dw}$ by $()^{'}$. Taking the
derivative is quite tedious, since $w$ appears in (\ref{eq:density-w
specific}) both in the coefficients of the Chebyshev polynomials and
in their argument. However, using elementary relations between the
Chebyshev polynomials and their derivatives, and after some lengthy
but straightforward computations, one gets:
\begin{eqnarray}
\label{eq:alll} a_l=(d-2)\left(
\frac{1}{4}(l-2)y_{l-2}(1)-\frac{1}{4}(l+2)y_{l+2}(1)-y_l(1)
\right)+\nonumber\\
(d-1)\left(
y^{'}_{l}(1)-\frac{1}{2}y^{'}_{l-2}(1)-\frac{1}{2}y^{'}_{l+2}(1)
\right)
\end{eqnarray}
We now recall that:
\begin{equation}
y_l(w) = \frac{1}{V}\frac{trY^l(w)-(d-w)^l}{(w(d-w))^{\frac{l}{2}}}
\end{equation}
and we write (\ref{eq:alll}) in terms of the $\tr Y^l$:
\begin{eqnarray}
\label{a_l in terms of Y} a_l=\frac{1}{2}\left(
(d-2)(l-2)\frac{trY^{l-2}(1)}{(d-1)^{\frac{l-2}{2}}}-(d-2)(l+2) \frac{trY^{l}(1)}{(d-1)^{\frac{l}{2}}}\right)\nonumber\\
+\frac{1}{2}\left(-\frac{(trY^{l-2})^{'}(1)}{(d-1)^{\frac{l-4}{2}}}
+2\frac{(trY^{l})^{'}(1)}{(d-1)^{\frac{l-2}{2}}}-\frac{(trY^{l+2})^{'}(1)}{(d-1)^{\frac{l}{2}}}
\right)
\end{eqnarray}
The requirement that  $a_l=0$, results in an expression for
$(trY^{l})^{'}$ in terms of $(trY^{k})$, where $k<l$. It is
convenient to define:
\begin{eqnarray}
  \label{p and q}
p_l \equiv \frac{(trY^{l})^{'}(1)}{(d-1)^{\frac{l-2}{2}}}\nonumber\\
q_l \equiv \left(
(d-2)(l-2)\frac{trY^{l-2}(1)}{(d-1)^{\frac{l-2}{2}}}-(d-2)(l+2)
\frac{trY^{l}(1)}{(d-1)^{\frac{l}{2}}}\right)
\end{eqnarray}
and after some further computations, the following inhomogeneous
recursion relation emerges:
\begin{eqnarray}
  \label{eq:periodic walks
with tails} \hspace{-2cm}
 p_{l+2}-p_l = \left \{
 \begin{array} {l l}  \sum_{k=1}^{l/2}q_{2k},
 & {\rm{if\ \textit{l}\ is\ even}}  \\
 \\
 \sum_{k=0}^{\lfloor l/2 \rfloor}q_{2k+1}, & {\rm if\ \textit{l}\ is\
odd}\
 \end{array}
 \right .
\end{eqnarray}
Substituting $q_l$ from(\ref{p and q}), we get that:
\begin{eqnarray}
  \label{eq:periodic walks
with tails} \hspace{-2cm}
 p_{l+2}-p_l = \left \{
 \begin{array} {l l} -2(d-2)\left[\sum_{k=1}^{l/2}\frac{trY^{2k}(1)}{(d-1)^k}+
\frac{l}{2}\frac{trY^{l}(1)}{(d-1)^{\frac{l}{2}}}\right],
 & {\rm{if\ \textit{l}\ is\ even}}  \\
 \\
 -2(d-2)\left[\sum_{k=1}^{\lfloor
l/2\rfloor}\frac{trY^{2k+1}(1)}{(d-1)^{k+\frac{1}{2}}}+
\frac{l}{2}\frac{trY^{l}(1)}{(d-1)^{\frac{l}{2}}}\right], & {\rm if\
\textit{l}\ is\ odd}\
 \end{array}
 \right .
\end{eqnarray}
It can be verified by substitution, that the solution of the
recursion relation is:
\begin{eqnarray}
  \label{eq:periodic walks
with tails} \hspace{-2cm} p_l = \left \{
 \begin{array} {l l} -l(d-2)\sum_{k=1}^{l/2-1}\frac{trY^{2k}(1)}{(d-1)^k}\ ,
 & {\rm{if\ \textit{l}\ is\ even}}  \\
 \\
  -l(d-2)\sum_{k=1}^{\lfloor
l/2\rfloor-1}\frac{trY^{2k+1}(1)}{(d-1)^{k+\frac{1}{2}}}\ , & {\rm
if\ \textit{l}\ is\ odd}\
 \end{array}
 \right .
\end{eqnarray}
or, equivalently:
\begin{eqnarray}
  \label{eq:periodic walks
with tails} \hspace{-2cm}
 (trY^{l})^{'}(1) = \left \{
 \begin{array} {l l}
 -l(d-2)\sum_{k=1}^{l/2-1}trY^{2k}(1)(d-1)^{\frac{l-2k-2}{2}},
 & {\rm{if\ \textit{l}\ is\ even}}  \\
 \\
 -l(d-2)\sum_{k=1}^{\lfloor
l/2\rfloor-1}trY^{2k+1}(1)(d-1)^{\frac{l-2k-3}{2}}, & {\rm if\
\textit{l}\ is\ odd}\
 \end{array}
 \right .
\end{eqnarray}
\begin{figure}[!h]
  \centering
%  \scalebox{0.5}{\includegraphics{tail.png}}
\scalebox{0.5}{\includegraphics{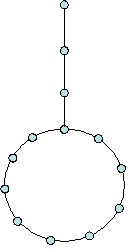}}
 \caption{A periodic walk of length $16$ composed of a cycle of length $10$ and a tail of length $6$}
 \label{fig:cycle with tail}
\end{figure}
Finally, $N(t;g=1)=-(trY^{l})^{'}(1)$ is the number of periodic
walks of length $l$ with one back-scatter. This result can be
derived directly by the following argument. For simplicity, let us
consider the case where $l$ is even. A periodic walk of length $l$
with one back-scatter must consist of a periodic walk of length $2k$
with no back-scatter, and a `tail' of length $l-2k$. A tail is a
periodic walk which goes over some path, and then comes back in the
opposite direction. Clearly, it has one back-scatter, as
demonstrated in figure (\ref{fig:cycle with tail}). The number of
periodic walks of length $2k$ with no back-scatter, is just
$trY^{2k}(1)$. The first edge in the tail can be chosen in $d-2$
ways. Any other edge, until we reach the end of the tail, can be
chosen in $d-1$ ways. There are $\frac{l-2k-2}{2}$ such choices to
make. The way back along the tail can be made in only one way. We
now need to sum over $k$. Since both $l$, and the length of the
tail, are even, the shortest periodic walk with no back-scatter is
of length $4$, so the sum must start at $k = 2$, and it must end at
$k = \frac{l}{2}-1$, since the shortest tail is of length $2$.

We have thus seen that using the $w$-trace formula, we were able to
extract combinatorial information about the graph. The computation
above was the simplest case, but it is clear that $N(t;g>1)$ can be
derived by further differentiations, and the use of known recursion
relations involving the Chebyshev polynomials and their derivatives.

\section{Unitary Evolution}
  \label{sec:unitary}
The trace formula (\ref{eq:traceformula}) which was discussed in
section (\ref{sec:intro}) was originally derived by constructing the
graph evolution operator $U(\mu)$ (\ref{eq:evolution}). The zeros of
the secular function $z(\mu)\doteq \det (I-U(\mu))$ provide the
spectrum of the graph, and using standard methods, the trace formula
followed. In view of the results derived in the previous sections,
it is natural to expect that together with $U(\mu)$, there exists a
one parameter family of \emph{unitary} operators, by which other
secular equations can be written down, and corresponding trace
formula can be derived. This is indeed the case, and to prove it, we
go back to the Bartholdi identity for regular graphs
(\ref{eq:bartholdi}). It is convenient to slightly modify the free
parameters $w,s$ and use $\alpha = s \ , \ \beta= w s$, Bartholdi's
formula reads now:
\begin{equation}
\hspace{-10mm}\det(I^{(2E)}-(\alpha B - \beta J)) =
(1-\beta^2)^{E-V} \det(( 1  + d\alpha \beta -
 \beta^{2} )I^{(V)} - \alpha A) \label{eq:bart}
\end{equation}
for any complex $\alpha\neq 0,\beta$.

Defining the $2E\times 2E$ matrix $U=\alpha B - \beta J$ and using
the properties of $B,J$ it is straightforward to show that
\begin{eqnarray}
U \ {\rm is \ unitary }\ \ \ \   \iff \ \ \ \ {\rm  both }\ \left\{
\begin{array} {l}
|\beta |^2 =1  \\
|\alpha |^2 d = \alpha \beta^* + \alpha^* \beta \
\label{unitaritycond}
\end{array}\right. \ .
\end{eqnarray}

The above implies that $\beta = {\rm e}^{i\phi}$ where $\phi$ is an
arbitrary real parameter.  If we impose further the relation
\begin{equation}
\alpha \mu =1 + d\alpha\beta -  \beta^{2} \   \label{def:mu}
\end{equation}
between $\alpha,\beta$ and the spectral parameter $\mu$, Bartholdi's
formula may be written as:
\begin{equation}
\hspace{-10mm}\det(I^{(2E)}-U(\mu ,\phi)) = (1-e^{2 i \phi})^{E}(\mu
- d e^{i \phi})^{-V}\det(\mu I^{(V)} - A) \label{eq:bartlam}
\end{equation}
where
\[U(\mu
,\phi)=\frac{1-e^{2 i \phi}}{\mu - d e^{i \phi}}B-e^{i \phi}J \] is
a unitary evolution operator depending parametrically on the real
and independent parameters $\mu$ and $\phi$ in the domain
$\mathcal{D}=\{ \mu \in{(-d,d)},\   \phi\in  \mathbb{R} \setminus
\{\pi \mathbb{Z}\}\}$. $\phi$ is restricted away from zero and any
integer multiples of $\pi$ since for these values $\alpha=0$ and
(\ref{eq:bartlam}) does not provide a relationship between the
spectrum of $A$ and $U(\mu,0)=-J$.

For  $\mu,\phi \in \mathcal{D}$ the secular equation
\[ Z_A (\mu)\ \doteq \ \det (\mu I-A)\ =\ (1-e^{2 i \phi})^{-E}(\mu - d e^{i \phi})^{V}
\det (I^{(2E)} - U(\mu ,\phi)) \] is the characteristic polynomial
of $A$ which is real on the real axis and its zeros coincide with
the spectrum of $A$. The spectral density function, $ \rho(\mu)
\equiv \frac{1}{V}\sum_{j=1}^V \delta(\mu - \mu_j) $, which can be
written as:
\begin{equation}
\rho(\mu) = -\frac{1}{V \pi} \lim_{\epsilon \rightarrow 0^+}
{\mathcal Im}\ \frac {{\rm d}\ }{{\rm d}\mu} \log Z_A(\mu +
i\epsilon)
\end{equation}
will now be a sum of a ``smooth" contribution from the phase of the
$(1-e^{2 i \phi})^{-E}(\mu - d e^{i \phi})^{V}$ term and a
fluctuating contribution from the sum over periodic orbits on the
graph, with amplitudes which are determined by the forward and
backward scattering defined by the evolution operator $U(\mu
,\phi)$:
\begin{eqnarray}
\rho(\mu) = & -\frac{1}{V\pi} \frac {{\rm d}\ }{{\rm d}\mu}
\log{((1-e^{2 i \phi})^{-E}(\mu - d e^{i \phi})^{V})} \nonumber \\
& + \ \frac{1}{V \pi}\lim_{\epsilon\rightarrow 0^+} {\mathcal Im}\
\frac {{\rm d}\ }{{\rm d}\mu} \sum_{t=1}^{\infty}\ \frac{1}{t}\tr
\left ( U(\mu+i\epsilon)^t\right) \label{eq:traceU}
\end{eqnarray}
Various real functions $\phi = \phi (\mu)$ can be defined, which
yield various smooth and fluctuating parts. One interesting case is
choosing $\phi = const$. For this choice the spectral density
function takes the form:
\begin{eqnarray}
\rho(\mu) = & -\frac{1}{\pi d} \ \frac{\sin{\phi}}{(\frac{\mu}{d})^2
+ 1 - 2 \frac{\mu}{d} \cos{\phi}} \\ & + \ \frac{1}{V
\pi}\lim_{\epsilon\rightarrow 0^+} {\mathcal Im}\ \frac {{\rm d}\
}{{\rm d}\mu} \sum_{t=1}^{\infty}\ \frac{1}{t}\tr \left (
U(\mu+i\epsilon)^t\right) \label{eq:const phi}
\end{eqnarray}
where
\begin{eqnarray}
\hspace{-10mm}\tr U(\mu)^t& = &\frac{(2 \sin{\phi})^t }{(\mu^2 + d^2
- 2\mu d \cos{\phi})^{\frac{t}{2}}}e^{i t (\phi-\arctan\frac{d
\cos{\phi}-\mu}{d \sin{\phi}})}\cdot \nonumber \\ & & \cdot\sum_g
N(g;t)
 \ \tilde{a} ^g \ e^{ig(\frac{\pi}{2}-\phi+\arctan\frac{d\sin{2\phi} - \sin{2\phi} -
\mu\sin{\phi}}{1-\cos{2\phi}-\mu \cos{\phi} + d\cos{2\phi}})}\
\end{eqnarray}
and
\begin{eqnarray}
 \tilde{a}=\tilde{a}(\mu)=\frac{\sqrt{2 + d(d-2) +\mu^2 - 2\mu d\cos{\phi} +
2(d-1)\cos{2\phi}}}{2 \sin{\phi}} . \nonumber
\end{eqnarray}

\begin{figure}[h]
  \centering
%  \scalebox{0.5}{\includegraphics{phikm.png}}
    \scalebox{0.5}{\includegraphics{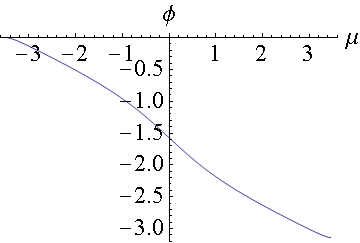}}
 \caption{The numerical solution of (\ref{eq:ode_KM}) for $d=4$ and $\frac{2k}{V}=-2$.}
 \label{figure:phi_KM}
\end{figure}

Specifically, the case of $\phi = -\frac{\pi}{2}$ corresponds to the
evolution operator chosen in (\ref{eq:traceformula}). Another
interesting choice would be that for which the smooth part is the
Kesten-McKay measure (\ref{eq:Mckay}). This will be achieved for
$\phi(\mu)$ which satisfies:
\begin{equation}
\frac{d}{2\pi}\cdot \frac{2\sin{\phi} - \frac{d\phi}{d\mu}[\mu^2
+d(d-2) - 2(d-1)\mu \cos{\phi}] }{\mu^2 + d^2 - 2\mu d \cos{\phi}} =
\rho_{{KM}}(\mu). \label{eq:ode_KM}
\end{equation}
where the L.H.S of this differential equation is the general
expression for the smooth part of the spectral density function,
normalized by the number of eigenvalues $V$, or equivalently
\begin{eqnarray}
\hspace{-10mm}\frac{2k}{V} + \frac{d}{2} \left(\frac{1}{2} -
\frac{\phi(\mu)}{\pi}\right) + \frac{\phi(\mu)}{\pi} \mp
 \nonumber \\ \hspace{-10mm}\mp\frac{1}{\pi}\arccos{\left(\frac{\mu \cos{\phi(\mu)} - d}
 {\sqrt{\mu^2+d^2-2\mu d\cos{\phi(\mu)}}}\right)}
-(-1\mp 1)=N_{KM}(\mu) \label{eq:NKM}
\end{eqnarray}
where the L.H.S is the general expression for the phase of $(1-e^{2
i \phi})^{-E}(\mu - d e^{i \phi})^{V} $ divided by $\pi V$,
$N_{KM}(\mu)=\int_{_{-2\sqrt{d-1}}}^{\mu}\rho_{{KM}}(\mu)$ is the
Kesten-McKay counting function and the $\mp$ refers to the sign of
$\mu\sin{\phi}$. The parameter $k\in{\mathbb{Z}}$ arises from the
fact that $(1-e^{2 i \phi})^{-E}(\mu - d e^{i \phi})^{V} $ is
defined up to an integer multiplicity of $2\pi$. Note that if
$\phi_{KM}^{(\frac{2k}{V})}(\mu)$ is the $k^{th}$ solution to
(\ref{eq:NKM}), then $\phi_{KM}^{(\frac{2k}{V})}(\mu)+ 2\pi$ is the
$k+\frac{(d-2)V}{2}$ solution, so in fact there are at most $
\frac{(d-2)V}{2}$ distinct solutions to (\ref{eq:ode_KM}) and
(\ref{eq:NKM}), where all the rest result by an addition of
multiples of $2\pi$. Figure (\ref{figure:phi_KM}) shows the
numerical solution of (\ref{eq:ode_KM}).\\
Assuming very large $d$ and scaling the spectral parameter,
$u=\frac{\mu}{2\sqrt{d-1}}$, one can expand in powers of
$\frac{1}{\sqrt{d-1}}$ so that (\ref{eq:ode_KM}) takes the form:
\begin{equation}
\frac{d\phi}{du}=-\frac{4}{d-1}\sqrt{1-u^2} +
\mathcal{O}((\frac{1}{d-1})^{\frac{3}{2}})\nonumber
\end{equation}
and solve to first order:
\begin{eqnarray}
\phi_{KM}^{(\frac{2k}{V})}(\mu)&=&C_k-\frac{1}{2(d-1)^2}\mu\sqrt{4(d-1)-\mu^2}
- \frac{2}{d-1}\arcsin{\frac{\mu}{2\sqrt{d-1}}} \nonumber \\
& & {\rm where} \ \ C_k=\frac{\pi}{2} +
\frac{2\pi}{d-2}(1+\frac{2k}{V}).
\end{eqnarray}

To every distinct solution corresponds a set of continuous operators
$U_{KM}^{(r)}(\mu)=U(\mu,\phi_{KM}^{(r)}(\mu))$ for which the smooth
part is the Kesten-McKay measure ($r \equiv\frac{2k}{V}$). Thus, the
fluctuating part must vanish after averaging over the ensemble, as
$V$ goes to infinity. Although a solution $\phi_{KM}^{(r)} $ can
cross the value $\phi_{KM}^{(r)}(\mu_0)=\pi k$ for $\mu_0\ne\pm d$,
hence is not a valid solution under the construction above, a
continuous set of unitary operators can still be defined using this
solution by:
\begin{eqnarray}
 U^{(r)}(\mu)=\left \{
 \begin{array} {l l}
 \frac{1-e^{2 i \phi_{KM}^{(r)}(\mu)}}{\mu - d e^{i \phi_{KM}^{(r)}(\mu)}}B-e^{i
 \phi_{KM}^{(r)}(\mu)}J & {\rm if}\ \mu\ne\mu_0 \\
 \\
 (-1)^k J & {\rm if}\ \mu=\mu_0
 \end{array}
 \right .
\end{eqnarray}
with the desired property.

\section{Discussion}
  \label{sec:discussion}
The novel feature which emerges from the present work is the fact
that the \emph{same} spectral density can be expressed by trace
formulae which are based on \emph{different} sets of periodic walks.
Moreover, the form of the smooth part and the weight associated to
periodic walks depend continuously on the parameters $w$ and $\phi$
in (\ref {eq:density-w specific}) and (\ref{eq:traceU}),
respectively. This is most prominent in the case $w=1$ where the
contributing walks are non back-scattering. For any $w \ne 1$,
back-scattering walks are contributing, so the weights and the
phases associated to each periodic walk are $w$ dependent.

The choice $w=1$ was beneficial for several reasons. First, it
serves as a new and independent proof of the Kesten-McKay formula
(\ref{eq:Mckay}), where the tree approximation is not taken
explicitly. Second, we were able to derive a leading order
correction for large, but finite, $V$.

The trace formulae we have derived, provide new insight into the
question raised in the introduction: Can we distinguish amongst the
periodic orbits on the graph distinct subsets, each responsible to a
different feature in the spectral density? As we have mentioned, the
smooth part (the Kesten-McKay measure) stems from periodic walks
restricted to tree-like subgraphs. We can now identify the
nb-periodic orbits as the subset which is responsible for the
fluctuating part of the spectral density. It is therefore
conjectured that for $w \ne 1$, the back-scattered walks give
opposite contributions to the smooth and oscillatory parts of
$\rho(\mu)$. This sheds a new light on the puzzle regarding the
re-summation which is required in (\ref{eq:traceformula}).

We have chosen to write the fluctuating part of the spectral
density, (\ref{eq:trace formula}) in a slightly peculiar way.
Usually in trace formulae, the summands in the sum over periodic
orbits, can be written as an amplitude times an exponent. The
exponent plays the r\^ole of classical action. We can thus identify
the quantity $\arccos{\frac{\mu}{2\sqrt{d-1}}}$ as an `action' per
unit step, of the quantum evolution on the graph. The amplitude is
proportional to $y_t$, which is a measure of the number of
t-periodic nb-orbits.

As we have shown, the $w$-trace formula can be used as a tool for
turning spectral information about graphs into combinatorial
information. This is one of the advantages of the $w$-trace formula
over the one specialized for $w=1$.

In the last section, we have proven that a set of unitary evolution
operators, which govern the quantum evolution on the graph, exist,
and that they reproduce the density of states for the discrete
Laplacian. This stands in contrast to the non-unitary ``evolution
operators" which were used in the preceding sections. As was shown,
a specific choice can be made, in order to reproduce the
Kesten-McKay measure (or the semi-circle distribution, for large
$d$).

In the following paper in the series, we shall use the fluctuating
part of (\ref{eq:trace formula}) in order to investigate the
spectral fluctuations of discrete Laplacians on regular graphs. We
shall show that these fluctuations are given by the appropriate
ensembles of Random Matrix Theory, and establish some interesting
connections between spectral and combinatorial graph theory.\\

\appendix

\section{The Bartholdi Identity}
 \label{sec:bartholdi}
The basis for the work presented in this paper, is an identity
proved recently by L. Bartholdi \cite{bartholdi} which reveals a
number of interesting connections between the spectral properties of
the graph and its set of periodic orbits. This identity  generalizes
a previous result by Bass \cite{bass}. For general graphs, the
Bartholdi identity is:
\begin{equation}
\hspace{-15mm}  \det(I^{(2E)}-s(B-wJ))=
 (1-w^2s^2)^{E-V}\det(I^{(V)}+ws^2(D-wI^{(V)})-sA)\ .
 \label{eq:general bartholdi}
 \end{equation}
The parameters $s$ and $w$ are arbitrary real or complex numbers,
$I^{(2E)}$ and $I^{(V)}$ are the identity matrices in dimensions
$2E$ and $V$, respectively, and the matrices $A,B,D$ and $J$ were
defined in the (\ref{sec:intro}). For $d$-regular graphs, the
Bartholdi identity reads:
\begin{equation}
\hspace{-15mm}  \det(I^{(2E)}-s(B-wJ))=
 (1-w^2s^2)^{E-V}\det(I^{(V)}(1+w(d-w)s^2)-sA)\ .
 \end{equation}
Its importance in the present context comes from the fact that it
connects the spectrum of the adjacency matrix $A$ with that of the
matrices $B$ and $J$, which can be used to count various types of
cycles and walks on the corresponding graph.

Bartholdi's  original proof of this identity is based on
combinatorial considerations. Proofs which are algebraic in nature
were provided by various authors, and we quote here the version
given by Mizuno and Sato \cite{MizunoSato2005} to make the present
paper self contained, and to provide a base for the derivation of
the trace formula for the discrete Laplacian.

\subsection{Proof of Bartholdi's identity}
The proof of Bartholdi's identity follows almost \emph{verbatim} the
proof presented in \cite{MizunoSato2005}.

Define the two $2E\times V$ rectangular matrices
\begin{eqnarray}
 B^{(+)}_{e,i} :=\left  \{
 \begin{array}{l l}
 1 & {\rm if}\ t(e)=i \\
 \\
 0 & {\rm otherwise}\ .
  \end{array}
 \right . \ \ \ ;\ \ \ B^{(-)}_{e,i} :=\left  \{
 \begin{array}{l l}
 1 & {\rm if}\ o(e)=i \\
 \\
 0 & {\rm otherwise}\ .
  \end{array}
 \right .
 \end{eqnarray}
 Denoting by $\widetilde{X}$ the transpose of $X$, one can easily prove
 that
 \begin{eqnarray}
& &B^{(+)} \widetilde{ B^{(-)}}  = B\ \ \ \ \  ;\ \
\widetilde{B^{(-)}} B^{(+)} = A\
\nonumber \\
 & & \widetilde{B^{(+)}} B^{(+)}  = d I^{(V)}\ ;\ \  B^{(+)}   \widetilde{B^{(+)}}   = Y
 J+I^{(2E)}\
 .
 \label{eq:identities}
 \end{eqnarray}
 Construct the two $(2E+V)\times (2E+V)$ square matrices
 \begin{eqnarray}
 \hspace{-20mm}L=\left[\begin{array}{l l}
 (1-w^2 s^2)I^{(V)} &  -  \widetilde{B^{(-)}} +w s \widetilde{B^{(+)}} \\
\\
 0 & I^{(2E)}
  \end{array}\right]
 \ ;
 \  M=\left[ \begin{array}{l l}
 I^{(V)} &  \widetilde{B^{(-)}} -w s \widetilde{B^{(+)}} \\
\\
 s B^{(+)} & (1-w^2s^2)I^{(2E)}
  \end{array}
 \right]
 \end{eqnarray}
 Using the identities (\ref{eq:identities}) one can compute the
 matrices $LM$ and $ML$, and since their determinants are equal, one
 finally gets the desired identity (\ref{eq:bartholdi}).

\subsection {Generalizations of the Bartholdi identity}
The Bartholdi identity can be easily generalized to three
interesting cases: ``Magnetic" graphs, multigraphs and weighted
graphs. The proof always follows the same steps shown above, and for
each case, one has to define the four matrices: $A,B,D,J$ properly.
\begin{itemize}
\item{Magnetic regular graphs:\\
We replace the adjacency matrix $A$ by its ``magnetic" analogue
(\ref{eq:magA}), and the edge connectivity matrix $B$ by
\begin{equation}
(B^{(M)})_{e,e'} = B_{e,e'} {\rm e}^{\frac{i}{2}(\phi_e+\phi_{e'})}
\label {eq:magB}
\end{equation}
The matrix $J$ is not modified since $\phi_e =-\phi_{\hat{e}}$. The
``magnetic" Bartholdi identity now reads:
\begin{equation}
 \hspace{-20mm} \det(I^{(2E)}-s(B^{(M)}-wJ))=
 (1-w^2s^2)^{E-V}\det(I^{(V)}(1+w(d-w)s^2)-sA^{(M)})\ .
 \label{eq:m-bartholdi}
 \end{equation}
The proof follows the same steps as above, after modifying
$B^{(\pm)}_{e,i}$ by multiplying them by ${\rm e}^{\pm
\frac{i}{2}\phi_e}$, and, by replacing the transpose operation ($\
\widetilde{\  }\ $) by hermitian conjugation. For non-regular
magnetic graphs, the matrix $D$ is the same as the non-magnetic
case.}
\item{Multigraphs:\\
The adjacency matrix is defined as explained in the introduction
(section (\ref{sec:intro})). $B$ is still a $(0,1)$ matrix where we
must list \textbf{all} the edges, including parallel ones and loops.
$J$ does not change and in $D$ we must count the degree of a vertex
including parallel edges and counting loops as two edges.}
\item{Weighted graphs:\\
The weighted adjacency matrix was defined in section
(\ref{sec:intro}). $J$ is not changed, and we redefine $B,D$ in the
following way:
\begin{equation}
(B^{(W)})_{e,e'} = B_{e,e'} \sqrt{W_e W_{e^{'}}} \label {eq:genB}
\end{equation}
\begin{equation}
(D^{(W)})_{i,j} =\delta_{ij}\sum_{e:t(e)=i} W_e \label {eq:genD}
\end{equation}}
\end{itemize}
These are three canonical generalizations. Obviously, one can make
further generalization by combining them (magnetic multigraph, for
example).

\section*{Acknowledgments}

\noindent The authors wish to express their gratitude to Mr. Sasha
Sodin for many insightful discussions, and for his assistance in
overcoming obstacles along the way. Prof. Nati Linial is also
acknowledged for explaining us some results from graph theory. We
are indebted to Dr I. Sato for reading the manuscript
and for several critical remarks and suggestions.\\
This work was supported by the Minerva Center for non-linear
Physics, the Einstein (Minerva) Center at the Weizmann Institute and
the Wales Institute of Mathematical and Computational Sciences)
(WIMCS). Grants from EPSRC (grant EP/G021287), BSF (grant 2006065)
and ISF (grant 166/09) are acknowledged.


\section*{References}
\begin{thebibliography}{10}
\bibitem {Imry book} Y. Imry, Introduction to Mesoscopic Physics, Oxford University Press, 1997.

\bibitem{percolation} Geoffrey Grimmett, Percolation, 2nd Edition,
Grundlehren der mathematischen Wissenschaften, vol 321, Springer,
1999.

\bibitem {probability} Geoffrey Grimmett and David Stirzaker,
Probability and Random Processes, Oxford University Press (2001).

\bibitem{Rudnick}  D. Jakobson, S. Miller, I. Rivin and Z. Rudnick. \emph{Level
spacings for regular graphs}, IMA Volumes in Mathematics and its
Applications 109 (1999), 317-329.

\bibitem {BGS}O.~Bohigas, M.-J. Giannoni, and C.~Schmit  \emph{Characterization of chaotic
  quantum spectra and universality of level fluctuation laws},  Phys.\
  Rev.\ Lett.  {\bf 52}, pp.~1--4  (1984) .

\bibitem{Gutzwiller} M.C. Gutzwiller, J. Math. Phys. {\bf 12} 343
(1971).

\bibitem{Berry} M.V. Berry, \emph{Semiclassical Theory of Spectral
Rigidity},  Proceedings of the Royal Society of London. Series A,
Mathematical and Physical Sciences, Vol. 400, No. 1819 (Aug. 8,
1985), pp. 229-251

\bibitem{Sieber} M. Sieber, K. Richter,\emph{ Correlations between Periodic Orbits and their R\^{o}le in Spectral
Statistics}, Physica Scripta, Volume T90, Issue 1, pp. 128-133.

\bibitem{Haake} S. Heusler, S. M¨uller, P. Braun, and F. Haake, U\emph{niversal spectral form
factor for chaotic dynamics}, J. Phys. A 37, L31 (2004).

\bibitem{hashimoto89}
Hashimoto K 1989 \emph{Zeta functions of finite graphs and
representations of  {$p$}-adic groups} {\em Automorphic forms and
geometry of arithmetic varieties\/} ({\em Adv. Stud. Pure Math.\/}
vol~15) (Boston, MA: Academic Press) pp 211--280

\bibitem{Avron} J.E. Avron, A. Raveh and B. Zur, \emph{Adiabtaic quantum
transport in multiply connected systems}, Reviews of Modern Physics,
Vol. 60, No. 4 (1988).

\bibitem{Kesten}H. Kesten
\emph{Symmetric random walks on groups}, Trans. Am. Math. Soc.
\textbf{92}, 336–354 (1959).

\bibitem{McKay}
McKay, B. D., \emph{The expected eigenvalue distribution of a random
labelled regular graph},  Linear Algebr. Appl. \textbf{40}, 203–216
(1981).

\bibitem{Brooks91} Robert Brooks,
\emph{The Spectral Geometry of k-Regular Graphs}, J. d'Analyse
\textbf{57}  120-151,(1991).

\bibitem{Alon} N. Alon, I. Benjamini, E. Lubetzky, S. Sodin,
\emph{Non-backtracking random walks mix faster},
arXiv:math/0610550v1

\bibitem{Harrison} J.M. Harrison, U. Smilansky and B. Winn, \emph{Quantum graphs where back-scattering
is prohibited}, J. Phys. A: Math. Theor. 40 (2007) 14181–14193

\bibitem{US07}Uzy Smilansky, Quantum Chaos on Discrete Graphs, J. Phys. A: Math. Theor. {\bf 40}
(2007) F621-F630.

\bibitem{KS} T. Kottos and U. Smilansky,
\emph{Quantum Chaos on Graphs}, Phys. Rev. Lett.  {\bf 79},4794-
4797, (1997). and  \emph{Periodic orbit theory and spectral
statistics for quantum  graphs}, Annals of Physics {\bf 274}, 76-124
(1999).


\bibitem{Roth} J. P. Roth in ``Lecture notes in Mathematics:
Theorie de Potential" (A. Dold and B. Eckmann, Eds.) p. 521,
Springer Verlag, New-York/Berlin (1985).


\bibitem {bartholdi} L. Bartholdi, \emph{Counting paths in graphs.}
Enseign. Math {\bf 45}, 83-131, (1999).

\bibitem {bass}
H. Bass, \emph{The Ihara -Selberg zeta function of a tree lattice},
Internat. J. Math. {\b 3}, 717-797 (1992).

\bibitem {MizunoSato2005}H. Mizuno and I. Sato, \emph{A new proof of Bartholdi's Theorem},
Journal of Algebraic Combinatorics, {\bf 22}, 259-271, (2005).

\bibitem{RamMurty} M. Ram Murty,
Ramanjuan Graphs, J. Ramanujan Math. Soc, {\bf 18} 1-20, (2003).


\bibitem{Mizuno08} H. Mizuno and I. Sato
\emph{The Scattering Matrix of a Graph},  The Electronic Jour. of
Combinatorics {\bf 15} R96, (2008).


\bibitem {Shnir} A.I. Shnirelman, \emph{Ergodic properties of
eigenfunctions}, Uspehi Mat. Nauk 29(6(180)), 181–182 (1974)

\bibitem {RBUS} R. Bl\"umel and U. Smilansky,
\emph{Random matrix description of chaotic scattering: Semi
Classical Approach.} Phys. Rev. Lett. {\bf 64}, 241--244 (1990).

\bibitem {Gutzbook} M. Gutzwiller  \emph{Chaos in Classical and Quantum Mechanics}
 Springer Verlag, New York, (1991).


\bibitem {Haakebook} Fritz Haake,
\emph{Quantum Signatures Of Chaos}. Springer-Verlag Berlin and
Heidelberg, (2001).

\bibitem {Stoekbook} H. J. Stoeckmann,  \emph{Quantum Chaos - An
Introduction}, Cambridge University press, Cambridge UK, (1990).

\bibitem{gnuzus} Sven Gnutzmann and Uzy Smilansky,
\emph{Quantum Graphs: Applications to Quantum Chaos and Universal
Spectral Statistics.} Advances in Physics {bf 55} (2006) 527-625.


\bibitem {Balobas}B. Bollobas, Random Graphs, Academic Press, London
(1985).

\bibitem{Chung}  Fan R. K. Chung, \emph{Spectral Graph Theory}, Regional
Conference Series in Mathematics  {\bf 92},  American mathematical
Society(1997).

\bibitem{AudreyBook} A. A. Terras, \emph{Fourier Analysis on Finite
Groups and Applications} London Mathematical Society Student Texts
{\bf 43} Cambridge University Press, Cambridge UK (1999).

\bibitem{AudreyRev} A. A. Terras, \emph{Arithmetic Quantum Chaos},
IAS/Park City Mathematical Series {\bf 12} 2002 333-375


\bibitem {Hurt}Hurt, N.E. \emph{The prime geodesic theorem and quantum
mechanics on finite volume graphs:a review.} Rev. Math. Phys. {\bf
13}, (2001), 1459-1503.

\bibitem{Ihara} Y. Ihara, \emph{On discrete subgroups of the two by two
projective linear group over a p-adic field}, J. Mat. Soc. Japan,
{\bf 18} (1966), 219-235.


\bibitem{Stark} H. M. Stark and A. A. Terras, \emph{Zeta Functions of
Finite Graphs and Coverings}, Adv. in Math. {\bf 121}, (1996)
124-165.

\bibitem{Sunada} Motoko Kotani and Toshikazu Sunada,\emph{ Zeta
Functions on Finite Graphs}, J. Math. Sci. Univ. Tokyo {\bf 7}
 no. 1, 7-25, (2000).

\bibitem{Starkmultipath} H.M. Stark, \emph{Multipath zeta functions of
graphs}, in: Emerging Applications of Number Theory, Minneapolis,
MN, 1996; IMA Vol. Math. Appl. 109 (1999) 601-615.

\bibitem{Janson} S. Janson, T. {\L}uczak and A. Ruci\'{n}ki \emph{Random
Graphs}, John Wiley \& Sons, Inc.

\bibitem{Mnev} P. Mn\"{e}v \emph{Discrete Path Integral Approach to the
Selberg Trace Formula for Regular Graphs}, Commun. Math. Phys. 274,
233-241 (2007).

\bibitem{Hoory}  S. Hoory, N. Linial, and A. Wigderson \emph{Expander Graphs and Their
Applications}, Bulletin (New Series) of the American Mathematical
Society {\bf 43}, Number 4, (2006), 439-561 S 0273-0979(06)01126-8

\bibitem{SashaSodin} S. Sodin, The Tracy-Widom law for some sparse random matrices,
arXiv:0903.4295v2 [math-ph] 7 Apr 2009.

\end{thebibliography}
\end{document}